\documentclass[
reprint,
superscriptaddress,
amsmath,
amssymb,
aps,
prl,
floatfix,
nobibnotes
]{revtex4-1}
\usepackage{hyperref}

\usepackage{amsmath}
\usepackage{amssymb}
\usepackage{bbold}
\usepackage[normalem]{ulem}
\usepackage{hyperref}
\usepackage{graphicx}
\usepackage{mathtools}
\usepackage{xcolor}
\usepackage{textcomp}
\usepackage{gensymb}
\usepackage{float}

\def\ute2{UTe$_2$}
\def\d2h{D$_{2h}$}

\begin{document}
\title{Field-angle-resolved heat transport in UTe$_2$: 
determination of nodal positions in the superconducting order parameter}


\author{Ian M. Hayes}
\affiliation{Maryland Quantum Materials Center, Department of Physics, University of Maryland, College Park, MD 20742, USA.}

\author{Elliot Fang}
\affiliation{Maryland Quantum Materials Center, Department of Physics, University of Maryland, College Park, MD 20742, USA.}

\author{Shanta R. Saha}
\affiliation{Maryland Quantum Materials Center, Department of Physics, University of Maryland, College Park, MD 20742, USA.}

\author{Vivek Mishra}

\author{P.J. Hirschfeld}
\affiliation{Department of Physics, University of Florida, Gainesville, FL, 32611-8440}

\author{Johnpierre Paglione}
\email{Corresponding Author: Mail adddress: paglione@umd.edu}
\affiliation{Maryland Quantum Materials Center, Department of Physics, University of Maryland, College Park, MD 20742, USA.}
\affiliation{The Canadian Institute for Advanced Research, Toronto, Ontario, Canada.}

\begin{abstract} 
One of the recurring hurdles in studying unconventional superconductivity is the challenge of efficiently and conclusively identifying the symmetry of the superconducting order parameter in a new material.
Uranium ditelluride (UTe$_2$) exhibits an unprecedented number of superconducting phases as a function of pressure and magnetic field, each presumably characterized by a different symmetry of the superconducting gap function. None of these phases has had its symmetry conclusively identified so far. In this article, we report results of an extensive study of the thermal conductivity of UTe$_2$ in its low-field, low-temperature superconducting state as a function of the angle of an applied magnetic field rotated in the $b$-$c$ plane. 
We observe clear and substantial oscillations in the thermal conductivity as a function of field angle, which naturally suggests the existence of point nodes in the gap.
Utilizing the experimentally determined Fermi surface, we are able to model this phenomenon for all the potential gap structures in UTe$_2$ and positively identify the location of these nodes as being along the crystallographic $b$-axis, implying that the superconducting order parameter most likely belongs to the $B_{2u}$ irreducible representation of the crystal point group. The clarity of this result will accelerate the identification of other superconducting phases in UTe$_2$, and guide future studies through the use of high resolution field-angle-dependent measurements.

\end{abstract}

\maketitle

\section{Introduction}

The heavy-fermion compound UTe$_2$ is a leading candidate for  intrinsic spin-triplet superconductivity \cite{ran_nearly_2019,Aoki2022,Lewin_2023} for several reasons, including the survival of the superconducting state to very large magnetic fields exceeding the Pauli limit and the weak temperature dependence of the Knight shift below the critical temperature \cite{ran_nearly_2019,Knebel2019,Nakamine2019,Aoki2019,Fujibayashi2022,Matsumura2023}. As one of  the few likely odd-parity superconductors known to exist, UTe$_2$ offers an extremely rare opportunity to empirically study the origin of triplet pairing. An important step in this project is to identify the full symmetry of the superconducting order parameter. The presence or absence of nodes in the superconducting energy gap and their position in momentum space reflect the symmetry of the interactions that drive the pairing, and are crucial to completing this task. 

UTe$_2$ forms in a body-centered orthorhombic crystal structure with $D_{2h}$ point group symmetry. In the presence of strong spin-orbit coupling this point group allows four possible one-dimensional irreducible representations (irreps) for a spin-triplet superconducting order parameter: $A_{u}$, $B_{1u}$, $B_{2u}$, and $B_{3u}$. 
At present, the nature of superconductivity in UTe$_2$ remains controversial, with evidence adduced in favor of all four allowed possible single-component irreps as well as proposals for non-unitary superpositions that break time reversal symmetry (TRS) ~\cite{Kittaka,jiao_microscopic_2019, Bae2021,Hayes2021, Wei2022, Ishihara2023,Iguchi2023,lee2023}. 
Penetration depth \cite{Ishihara2023,CarltonJones2025}, STM \cite{jiao_microscopic_2019}, and other experiments \cite{Hayes2021} have found signatures of a multi-component order parameter, while 
more recent tests of this scenario appear more consistent with a single-component (1D) state \cite{Theuss2024, Rosa2022, Ajeesh2023}. 
NMR measurements have shown finite changes in the Knight shift through $T_c$ for fields oriented along all crystal axes \cite{Nakamine2021,Matsumura2023}, implying the presence of a strong gap in all directions as expected for the nodeless A$_u$ state, also proposed by a previous thermal conductivity study \cite{suetsugu2023}.
However, the majority of experiments have provided evidence in favor of low-energy excitations due to nodes in the superconducting gap. Thermodynamic measurements, including heat capacity \cite{Kittaka,lee2023}, penetration depth \cite{Ishihara2023,CarltonJones2025}, ultrasound \cite{Theuss2024}, and thermal conductivity \cite{metz_point-node_2019,Hayes2025}, are all most consistent with nodes lying in the $ab$-plane. 
On the other hand, STM quasiparticle interference measurements were successfully modeled with a $B_{3u}$ ($a$-axis nodes) order parameter \cite{ShuqiuWang2025} and planar tunneling experiments have suggested a $B_{1u}$ (i.e., $c$-axis nodes) irrep. \cite{Li_Maple_PNAS2025}. 
Convergence on this question has been difficult due to the specific vulnerabilities of each measurement. The tunneling, NMR, and penetration depth experiments are all, to different extents, surface-sensitive. Ruling out the existence of distinctive physics or complicating factors on the surface is difficult, especially in UTe$_2$ where topological surface states are a real possibility \cite{jiao_microscopic_2019,Bae2021,Gu_Davis2025}. On the other hand, the measurements of bulk thermodynamic quantities typically cannot provide momentum-resolved information about the gap structure or anisotropy, meaning conclusions about the locations of nodes may be model-dependent.

\begin{figure*}
\includegraphics[width=15cm]%
{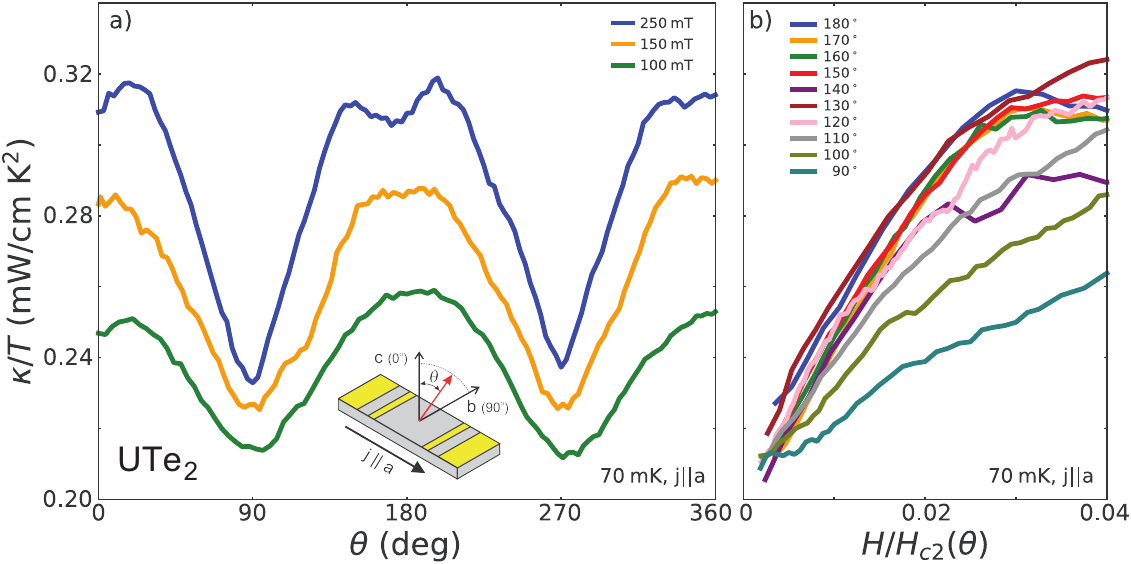}
\caption{
{\bf Magnetic field-angle dependence of thermal conductivity deep in the superconducting state of UTe$_2$.} 
Panel a) presents the angle-sweep data measured at a fixed temperature of 70~mK ($\sim 0.03 T_c$), with heat current applied parallel to the $a$-axis and field rotations through the perpendicular $b$-$c$ plane at fixed fields of 100, 150 and 250 mT. 
Panel (b) shows fixed-angle field sweeps with the same geometric configuration at the same temperature. The field values in this plot have been normalized to the angle-dependent upper critical field $H_{c2}(\theta)$ \cite{ran_extreme_2019}. 
}
\label{fig:angle_variation} 
\end{figure*}

Both of these limitations would be overcome by a bulk technique that can probe the quasiparticle spectrum in an angle-dependent way. In the so-called Volovik regime \cite{Volovik1993}, an applied magnetic field provides a way of shifting the quasiparticle spectrum, which can be observed through measurements of the heat capacity or thermal conductivity. As a result, thermal measurements under a continuously rotated magnetic field provide relatively direct and robust access to the k-space variation in the superconducting gap \cite{Matsuda2006}.

In the simplest picture, this tunability of the quasiparticle spectrum comes from the Doppler shift that is due to the superfluid flow around the vorticies that appear in the presence of an applied field. This induces a shift in the quasiparticle energy ${\bf p}_s\cdot {\bf v_F}$ that depends only on the local superflow momentum ${\bf p}_s({\bf r})$ and the Fermi velocity ${\bf v_F}$.  At low energies, quasiparticles are confined to the neighborhood of the nodes, ${\bf k}\sim {\bf k}_n$.  Since the supercurrent flows perpendicular to a vortex, such that the applied field ${\bf H}\perp {\bf p}_s$,  the largest Doppler shifts are obtained for ${\bf p}_s\parallel {\bf k}_n$, and therefore ${\bf H}\perp {\bf k}_n$. The largest field-dependent shift in the density of states, $\delta N({\bf H})=  N({\bf H})-N(0) $, is thus obtained when this condition is obeyed, provided all nodes have the same symmetry with respect to the field.  Thus the field angle-dependent density of states is expected to oscillate with field direction, and have a minimum when the field points along the nodal directions, driving a similar behavior in the Sommerfeld coefficient $\gamma(T,{\bf H})$.  With some caveats \cite{Matsuda2006}, the same argument applies to the electronic thermal conductivity $\kappa_{el}$, which can typically be measured to lower temperatures than specific heat.  Like the specific heat, the thermal conductivity field angle oscillations in a nodal superconductor are known to {\it invert} above a certain temperature, i.e. minima and maxima exchange roles \cite{Vekhter2006,Boyd2009}.

\begin{figure}
\includegraphics[width=1\columnwidth]{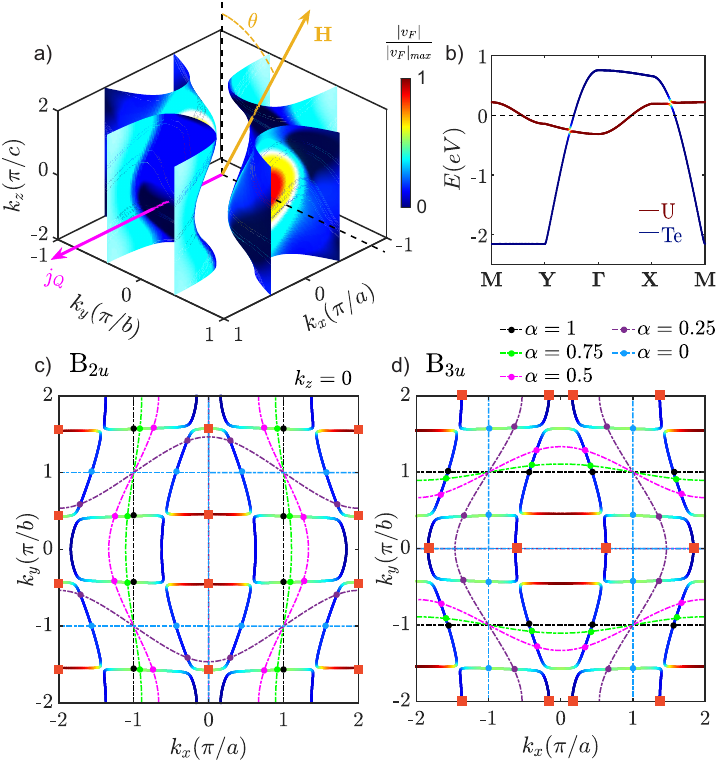} 
\caption{
{\bf Model Fermi surface of UTe$_2$ and gap node arrangements for thermal conductivity calculations.} 
Panel a) presents the three-dimensional Fermi surfaces of UTe$_2$ used to model thermal conductivity, where the color function depicts the the magnitude of the Fermi velocity normalized to its maximum value.  Heat current ($\parallel a$-axis) and magnetic field rotation ($b$-$c$) plane are indicated by magenta and yellow arrows, respectively.
Panel b) shows the dispersion of U and Te bands that cross the Fermi level along the high symmetry directions.
Panels c) and d) present the $k_z$=0 Fermi surface cuts and nodal positions for the possible $B_{2u}$ and $B_{3u}$ order parameter states, respectively, including symmetry-imposed nodes (red squares) and accidental nodes (circles) for various values of tuning parameter $\alpha$ which controls the positions of the accidental nodes (see text). The solid lines represent the Fermi surface contours, and the dotted-dashed lines indicate the zeros of the gap function.
}
\label{fig:fig2new} 
\end{figure}

Although this approach to studying the superconducting gap structure is powerful in theory, it has proven difficult to implement in practice \cite{Matsuda2006}. In order to achieve a meaningful variation of $\kappa/T$ as the field-angle is swept, the quasiparticle heat transport has to dominate while the thermally excited quasiparticles remain limited to the nodal regions. This necessitates measurements on very clean samples at very low temperatures, along with the experimental capability to resolve angular dependence at very low magnetic fields.
Here, we present field angle-dependent thermal conductivity measurements on a high quality crystal of UTe$_2$ (RRR = 600, T$_c$ = 2.1K, grown by the tellurium-flux method), probing the quasiparticle response very deep in the superconducting state. This sample was studied previously in \cite{Hayes2025} (sample S3 in that text), where we show the resistivity and thermal conductivity as a function of temperature in zero field. The measurements reported here were performed at temperatures well below $T_c$ ($\sim$ 2\% of $T_c$) and magnetic fields well below $H_{c2}$ ($\sim$ 1\% of $H_{c2}$), providing the most robust and accurate picture of gap structure to date.
We apply heat current along the crystallographic $a$-axis and rotate a transverse field through the $b$-$c$ plane of the crystal to search for signatures of nodes, finding large, distinct oscillations with minima along the $b$-axis. 
We compare our data with model calculations of the field-angle response of the thermal conductivity using a realistic tight-binding Fermi surface, finding agreement in the angular response of a model with symmetry-imposed nodes along the $b$ direction. We also examine the two component states which break TRS, leading to the splitting of the quasiparticle energy spectrum. The lower branch of the quasiparticle spectrum exhibits spectral  nodes  which are distinct from order parameter nodes, but function like usual nodes as they host low energy excitations. We find that, among  these states, only the states with spectral nodes along or very close to the $b$-axis are consistent with the experimental data. (see Supplementary Information for details). Lacking unequivocal evidence in favor of TRS breaking, we conclude that the zero-field, zero-pressure bulk superconducting order parameter of UTe$_2$ most likely has $B_{2u}$ symmetry.

\section{Results}

\begin{figure*}
\includegraphics[width=1\linewidth]{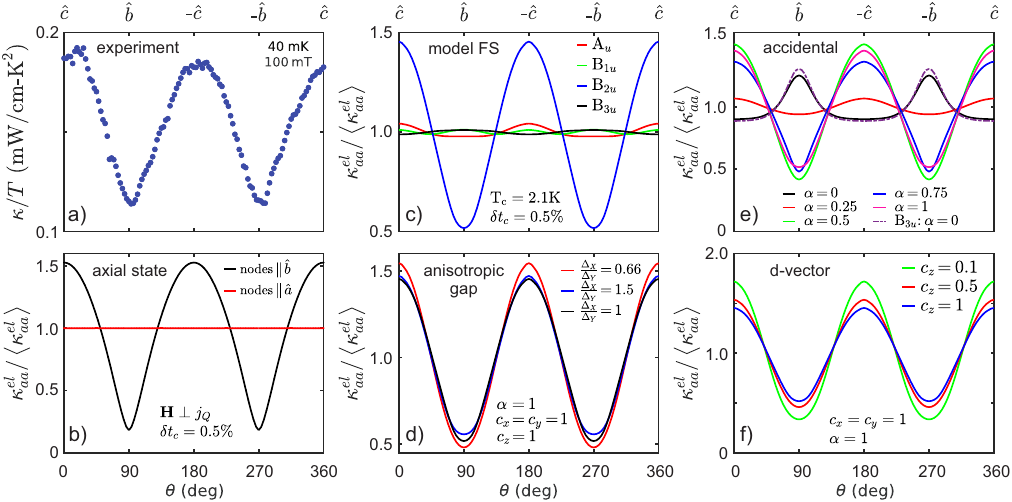} 
\caption{
{\bf Comparison of experimental $b$-$c$ plane field-angle sweep thermal conductivity to theoretical models.} 
Panel (a) presents experimental data measured at 40~mK for $a$-axis heat current and $b$-$c$ plane rotation of 100~mT field. Panels b)-f) present theoretical models of the angle-dependent electronic thermal conductivity normalized to its average 
for the same current-field orientation, for the following cases:  
b) for a simple axial state on a spherical Fermi surface  with linear point nodes positioned along the ${b}$ (black) and  ${a}$ (red) axes; 
c) for all four irreducible representations of the  D$_{\mathrm{2h}}$ states on realistic Fermi surfaces with equal magnitude gaps on two bands; 
d) for the $B_{2u}$ state with different gaps on two bands; 
e) for the $B_{2u}$ state with accidental nodes generated by parameter $\alpha$ that mixes nearest neighbor and next-nearest neighbor pairing ($B_{3u}$ state with only next-nearest neighbor pairing shown with dashed line); 
f) for the $B_{2u}$ state with variation of the coefficient of the $\hat{c}$ component of the $\mathbf{d}$-vector basis function. All theoretical calculations are done at 40~mK and 100~mT field. The impurity parameters are chosen to give $0.5\%$ $T_c$ suppression.
}
\label{fig:fig3new} 
\end{figure*}

Fig.~\ref{fig:angle_variation} presents $\kappa(H, \theta)/T$ data measured at 70~mK as functions of both field angle (fixed magnitude) and field magnitude (fixed angle), showing consistency between the measurements. Measurements up to 250~mT show clear two-fold angle oscillations with a substantial modulation of $\sim$25\% of the average value that increases on further cooling (c.f. 40~mK data in Fig.~ \ref{fig:fig3new}a). 
These oscillations are systematic as a function of field in the low-field range where we observe a quasi-linear growth in $\kappa/T$ for all angles in the $b$-$c$ plane (see Fig \ref{fig:angle_variation} (b)), and indicate the presence of strong angular variation in the gap.
The quasi-linear increase in $\kappa$ with field (also observed for the $H \parallel a$ orientation \cite{Hayes2025}) 
is most naturally consistent with the presence of Doppler-shifted nodal excitations due to the Volovik effect \cite{Volovik1993}. An alternative scenario in which the gap possess deep minima instead of nodes is strongly disfavored by the fact that there is no sign of activation in field or temperature. The field dependence is linear from the lowest fields we measured (see Fig  \ref{fig:angle_variation} (a)) and the slope decreases with increasing field. Additionally, the amplitude of the oscillations continues to increase even below 70 mK (cf. Fig. \ref{fig:fig3new} (a)), which could only be accommodated by assuming a gap anisotropy of much more than 10. Finally, these observations are consistent with our previous work which also failed to find evidence of activation with a field applied along the $a-$axis.
Second, the oscillations in $\kappa$ at fixed field do not simply follow the variation of $H_{c2}(\theta)$ in the plane of rotation. As shown in Fig.~\ref{fig:angle_variation}(b), normalizing $\kappa(H)$ to $H_{c2}(\theta)$ shows that an angular dependence much stronger than any upper critical field variation (weak in the b-c plane) is evident, ruling out any trivial explanation consistent with an isotropic gap.
Above 200~mT, one can observe that the field-angle dependence of $\kappa$ becomes more complex, suggesting that physics beyond the simple Doppler-shift picture becomes relevant. This begins to appear in the 250~mT field-angle sweep (Figure \ref{fig:angle_variation} panel (a)), as well as in the field sweep data (panel (b)) and in higher field angle sweeps shown in the Supplementary Information.  As a result, we limit our theoretical analysis to modeling the lowest-field measurements, well within the quasi-linear regime.

With minima and maxima in $\kappa(\theta)/T$ along the $b$ and $c$ axes respectively, this oscillation is exactly that expected for a system with point nodes in the gap \cite{Matsuda2006}. Since the Doppler shift is strongest when the field is perpendicular to the nodes and weakest when the field is along the nodes, the oscillatory pattern that we observe is most naturally consistent with a pair of nodes located along the crystal $b$ axis. However, the exact nature of the electronic structure can complicate these conclusions. In addition to the nodal contributions, it is well known that the angular variation of the Fermi velocity $\bf v_F$ also contributes to the magnetic field angle dependence of $\kappa({\bf H})$, and at significant temperatures and fields can overcompensate the nodal quasiparticle contribution.  Thus it is important in any theoretical analysis to properly model the Fermi surface so as to account for significant variations in $\bf v_F$.  Fortunately, accurate measurements of the Fermi surface of UTe$_2$ exist \cite{Fujimori2019,Miao2020,Aoki2022,Broyles2023,Eaton2024},  allowing the construction of appropriate tight-binding models reflecting this anisotropy \cite{Theuss2024,ShuqiuWang2025}, so that the nodal contributions can indeed be isolated.

We adopt the tight-binding model introduced by Theuss {\it et al.} \cite{Theuss2024}, which consists of two U and two Te bands and is qualitatively consistent with angle-resolved photoemission spectroscopy (ARPES) and quantum oscillation measurements \cite{Fujimori2019,Miao2020, Aoki2022,Broyles2023,Eaton2024}. This model comprises the U 6$d$  and Te 5$p$ bands that disperse along the $k_x$ and $k_y $ directions. The Hamiltonian for the two U/Te atoms is
\begin{eqnarray}
\mathcal{H}_{\mathrm{U/Te}} &=& \sum{\mathbf{k},\sigma}\Psi^\dagger_{U/Te} \begin{bmatrix}
{H}_{\mathrm{U/Te,  11}} & {H}_{\mathrm{U/Te, 12}} \\
{H}_{\mathrm{U/Te, 12}}^\ast & {H}_{\mathrm{U/Te,  11}}
\end{bmatrix} \Psi_{U/Te}\\
{H}_{\mathrm{U,11}}&=&\mu_\mathrm{U} - 2 t_\mathrm{U} \cos (k_x a) - 2t_\mathrm{ch,U} \cos (k_y b), \\
  {H}_{\mathrm{U, 12}}&=&-\Delta_\mathrm{U} - 2t'_\mathrm{U}  \cos (k_x a) -2t'_\mathrm{ch,U} \cos(k_y b) \nonumber \\
  & & - 4t_\mathrm{z,U} \exp(-ik_z c/2)\cos (k_x a/2) \cos (k_y b/2), \\ 
 {H}_{\mathrm{Te,11}}&=&\mu_\mathrm{Te}-2t_\mathrm{ch,Te}\cos (k_x a),\\  
 {H}_{\mathrm{Te,12}}&=&-\Delta_{Te}-t_\mathrm{Te} \exp(-i k_y b) \nonumber \\
 & & - 2t_\mathrm{z,Te} \cos (k_x a/2) \cos(k_y b/2) \cos (k_z c/2). 
\end{eqnarray}
Here $\Psi^{\dagger}_{U/Te}\equiv (c^\dagger_{U1/Te1,\mathbf{k}\sigma} c^\dagger_{U2/Te2,\mathbf{k}\sigma})$ is the basis composed of U/Te orbitals, $c^\dagger_{\mathbf{k}\sigma}/c_{\mathbf{k}\sigma}$ denotes the creation/annihilation operator for a fermion with momentum $\mathbf{k}$ and spin $\sigma$ in the respective orbital. This tight binding model results in two electronic bands that cross the Fermi level—one derived from the 'U bands' and one from the 'Te bands' (see Fig.~\ref{fig:fig2new}b)). These bands hybridize via a momentum-independent interaction ($\delta=$0.13eV) to form two distinct cylindrical bands 
\cite{ARPESnote}
: one with electron-like character and the other with hole-like character. The resulting Fermi surfaces are illustrated in Fig. \ref{fig:fig2new}a), where the color scale indicates the absolute value of the Fermi velocity, normalized to its maximum value. Key features emerge from this model, whose parameters are taken from Ref. \onlinecite{ShuqiuWang2025}: the Fermi velocity is notably low along the {U-dominated segments}, which in turn leads to a higher density of states in these regions.  We further scale the bands to match the experimentally measured normal state Sommerfeld coefficient of 120~mJK$^{-2}$mol$^{-1}$ \cite{ran_nearly_2019}.

Except where noted, the calculation of the thermal conductivity proceeds largely as in Refs. \cite{MishraHirschfeld2023,Hayes2025}, focusing on the electronic thermal conductivity $\kappa_{el}$.  
In the semi-classical approach\cite{Volovik1993,Kuebert1998,Matsuda2006}, the magnetic field Doppler shifts the energy in the superflow field of the vortex by an amount
\begin{equation}
\delta \omega = \frac{1}{2} m^\ast \mathbf{v}_F\cdot \mathbf{v}_s,
\label{eq:Dopplershift}
\end{equation}
where $\mathbf{v}_s$ is the superfluid velocity, $\mathbf{v}_F$ is the Fermi velocity, and $m^\ast$ is the effective mass.
In the isolated vortex approximation, valid for sufficiently weak fields in a strongly type-II superconductor, the superflow varies as 
\begin{eqnarray}
\mathbf{v}_s = \frac{\hbar}{2 m^\ast r}\hat{\Psi},
\end{eqnarray}
with  the distance from the vortex center $r$, where $\Psi$ is the winding angle.  The true anisotropy of the crystal is thus incorporated only into the momentum-dependent Fermi velocity in Eq. \eqref{eq:Dopplershift}.

We first present simple results for a spherical Fermi surface and a simple axial state as employed in Ref. \cite{Hayes2025}.  As in the experimental protocol described above, we assume the heat current is directed along the ${a}$ axis, and rotate the magnetic field in the $b$-$c$ plane.  Choosing the linear point nodes of the axial state to lie in either the ${a}$ or the ${b}$ directions then mimics the $B_{3u}$ and $B_{2u}$ states at the same level of approximation as in Ref. \onlinecite{Hayes2025}.   If the nodes are along the direction of the heat current, the absolute magnitude of $\kappa_{el}$ is larger than if the heat current direction were away from the nodes. As is clear from Fig. \ref{fig:fig3new}(b), however, there are no thermal conductivity oscillations since the Fermi velocity is assumed to be isotropic, and the nodes are always perpendicular to the field when it is rotated in the $b$-$c$ plane.  On the other hand, if the nodes are along ${b}$, while the heat current is smaller in magnitude, the Doppler shift varies significantly over the angular range of the experiment, leading to large, ${\cal O}$(1) oscillations in the normalized electronic heat conductivity, with a minimum as expected when the field is along the nodal direction.  These results compare favorably with the data shown in Fig. \ref{fig:fig3new}(a), but need to be revisited using more realistic models of the band structure and pairing states.
\begin{table}
\begin{tabular}{lcc}
\hline \hline Irreps: & $\mathbf{d}$-vector & Nodal positions \\
\hline$A_u$ & $(c_x X, c_y Y, c_z Z)$ & none \\
$B_{1 u}$ & $(c_x Y, c_y X, c_z W)$ & none \\
$B_{2 u}$ & $(c_x Z, c_y W, c_z X)$ & $b$-axis \\
$B_{3 u}$ & $(c_x W, c_y Z, c_z Y)$ & $a$-axis \\
\hline \hline
\end{tabular}
\caption{Forms of the  $\mathbf{d}$-vector for each symmetry allowed odd-parity irreducible representation (Irreps) of the $D_{2h}$ point group in the presence of strong spin-orbit coupling and the corresponding positions of symmetry imposed nodes on cylindrical Fermi surfaces open along $c$-axis. For a body centered orthorhombic crystal the basis functions are:  $X=\alpha\sin(k_x a)+(1-\alpha)\sin\left(k_x a/2\right)\cos\left(k_y b/2 \right)\cos\left(k_z c/2 \right)$, $Y=\alpha\sin(k_y b)+(1-\alpha)$ $\sin\left(k_y b/2 \right)\cos\left(k_x a/2\right)\cos\left(k_z c/2 \right)$, $Z=\sin\left(k_z c/2 \right)\cos\left(k_x a/2 \right)\cos\left(k_y b/2\right)$ and $W=\sin\left(k_x a/2 \right)$ $\sin\left(k_y b/2 \right)\sin\left(k_z c/2\right)$.   }
\label{tab:irreps}
\end{table}

To this end, we use the electronic structure model described above, with low-energy bands and Fermi surface shown in Fig. \ref{fig:fig2new}a), and compare first with results obtained for single-irreducible representation states.  As basis functions, we adopt the set generated for the $A_{u}$, $B_{1u}$, $B_{2u}$ and $B_{3u}$ irreps 
allowing first nearest-neighbor and second nearest-neighbor pairing terms \cite{Christiansen2025A,Tei2023}, with the periodicity of the body centered orthorhombic Brillouin zone. These functions have the symmetry of those in Table \ref{tab:irreps}, where $X,Y$ and $Z$ behave as $k_x,k_y$ and $k_z$ near the $\Gamma$ point, but have more complicated forms elsewhere in the Brillouin zone. As discussed in Ref.~\cite{Christiansen2025A}, in addition to the symmetry enforced nodes for each irrep ($\parallel a$ for $B_{3u}$, $\parallel b$ for $B_{2u}$, fully gapped for $\mathrm{A_{1u}}$ and $B_{1u}$), accidental nodes occur generically for these states over the given Fermi surface. These additional nodes,  together with their associated Fermi velocities, can contribute significantly to $\kappa_{el}$ at low temperatures.  Their locations are shown in Fig. \ref{fig:fig2new}(b) and (c) for various values of $\alpha$, a parameter controlling the relative weight of nearest- and next-nearest-neighbor pairing.

In Fig. \ref{fig:fig3new}(c)-(f), we now show plots of the predicted oscillations of $\kappa_{el}$ for field rotating in the $b$-$c$ plane under various assumptions regarding the parameters of the model.  The simplest choice is to replace  $X$ and $Y$ with $\sin (k_x a)$ and $\sin (k_y b)$ basis functions, which manifestly have the correct symmetry near $\Gamma$ point and correspond to next-nearest neighbor pairing in the $ab$-plane with $\alpha=1$ \cite{Christiansen2025A}. {The additional nodes in this case lie along segments of the Fermi surface (see crossings of black lines with the Fermi surface in Fig. \ref{fig:fig2new}(b)) that correspond to Fermi velocities pointing along the same directions as the symmetry-enforced nodes, so they will not significantly alter the form of the oscillations. } In the absence of experimental information about the relative sizes of the gaps on the two bands, we have further assumed them equal in the current analysis. Here one sees that, as expected, the $B_{2u}$ state produces at low $T$ the deep minima in the oscillations in $\kappa_{el}(\theta)$ with field angle along $b$, as in the spherical model shown in panel (b), while the oscillations for the other three irreps are negligible.

To test the robustness of this fit to the low-${T}$ data, and to rule out, if possible, other irreps, in Fig. \ref{fig:fig3new}(d)-(f) we have varied other parameters of the model.  In Panel (d), we show that changing the relative sizes of the two gaps does not change the qualitative form of the oscillations in the $B_{2u}$ case, for generic other parameters.  Further such variations are shown in the Supplementary Information.  In panel (e),  we illustrate how varying $\alpha$ {\it can} reverse the oscillations, since e.g. in the $B_{2u}$ case smaller $\alpha$ moves the accidental nodes towards the $a$ direction, which undermines the contribution of the  symmetry-enforced nodes due to the higher heat current when $j_Q\parallel $ nodes. The same effect does {\it not} occur, however, in the $B_{3u}$ case, since although the accidental nodes are along $b$, the heat current that they contribute there is dominated by the symmetry enforced nodes because $j_Q\parallel a$; the calculated oscillations in this case for $B_{3u}$ are also shown in panel (e), but they exhibit maxima rather than minima along $b$. Consequently, the occurrence of dominant nearest-neighbor pairing is improbable.   Finally, in (f) we investigate changes in the weight factors for the $X,Y$ and $Z$ basis functions, and show that this does not change the qualitative behavior.

\begin{figure}
\includegraphics[width=0.95\columnwidth]
{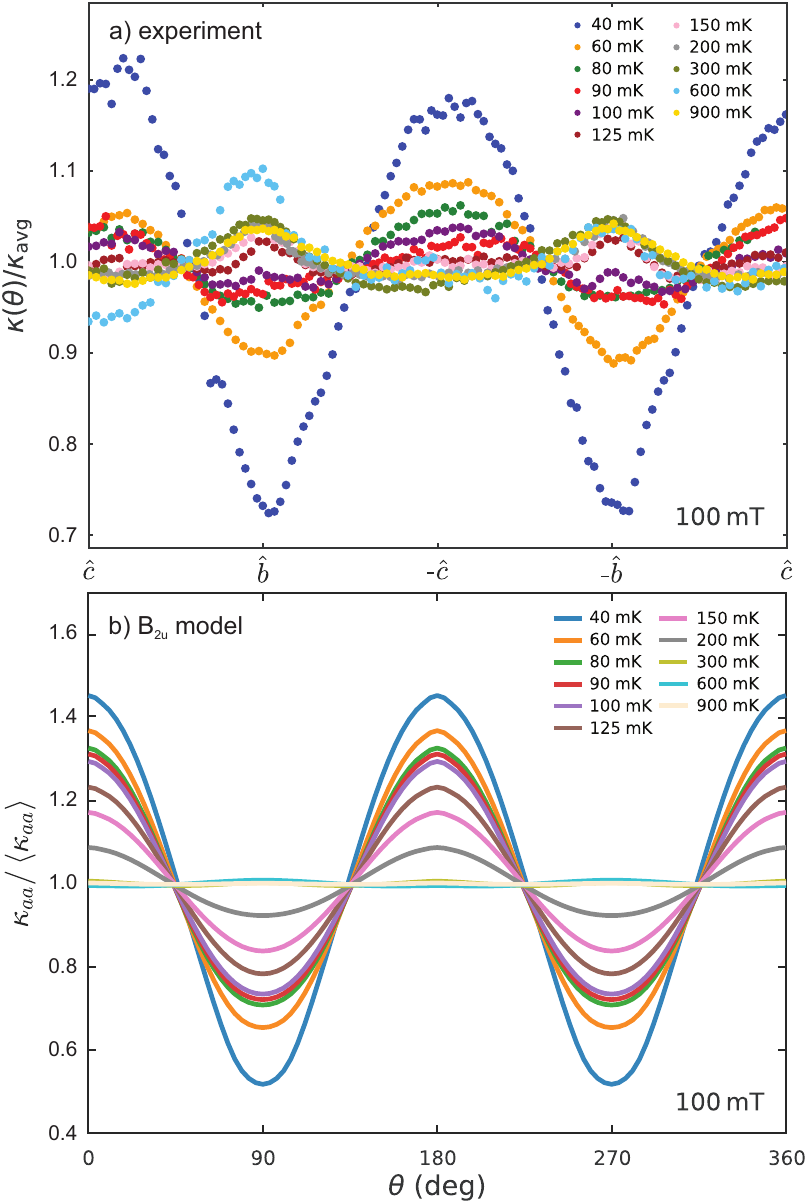}
\caption{
{\bf Temperature evolution of field-angle sweep $a$-axis thermal conductivity.} 
Panel (a) plots experimental data measured at fixed temperatures ranging from 40-900~mK and $b$-$c$ plane rotation of 100~mT field. Panel (b) presents the calculated angle-dependent thermal conductivity for the $B_{2u}$ state for 100~mT applied field and various temperatures. The impurity parameter gives $0.5\%$ suppression of the transition temperature with equal gap magnitudes on two bands. The coefficients of the basis functions are set to unity and only nearest neighbor pairing is included (\textit{i.e.,} $\alpha=1$).  
}
\label{fig:temperature_evolution} 
\end{figure}

Although the low-temperature regime possesses the fewest complicating factors, the behavior of $\kappa/T$ at higher temperatures can provide another important check on this conclusion. These oscillations in the thermal conductivity are expected to exhibit a characteristic inversion with $T$ predicted by both Brandt-Pesch-Tewordt theory \cite{Vekhter2006} and by the Doppler shift approach \cite{Boyd2009}, whereby the positions of the minima in $\kappa_{el}(T,H,\theta)$ shift by 90$^\circ$.  This occurs at a temperature scale that is difficult to express in simple terms but depends sensitively on $H$, impurity concentration, etc.  
Such an inversion is clearly manifested in the data (see Fig. \ref{fig:temperature_evolution}(a)). Experimentally, this occurs around 125~mK, whereas in the current theory for the $B_{2u}$ state it occurs at somewhat higher temperatures, about 300~mK, and is smaller in relative magnitude.  Nonetheless, given the many uncertainties in the electronic structure model and the pairing Ansatz, the fact that both the model and the data show an inversion is striking, and we have not attempted further fine tuning to achieve precise agreement on the temperature of inversion.
Notably, our models for other pairing symmetries such as $B_{3u}$ show significant departures from the data (see SI).

The question of the temperature dependence of $\kappa(H,\theta)/T$ raises the important issue of the phonon conductivity, which is expected to contribute a field- (and therefore field angle-) independent component to the measured values that should be considered in quantitative comparisons.
The expected zero-field power law temperature dependence of $\kappa_{el}(T)/T$ for a system with point nodes is $\sim T^2$ for $J_Q\parallel$ nodes, and $\sim T^4$ for $J_Q\perp$ nodes. This is markedly different than the experimentally measured quasi-linear temperature dependence of $\kappa(T)/T$ (see Supplementary Figure
\ref{fig:SI_Theory_3}), leading to the previous conclusion that the electronic term is intrinsically small and a relatively large phonon contribution explained the temperature dependence up to $T_c$ rather well \cite{Hayes2025}. 
This would imply that our ``fits" to the oscillation data presented in Figs. \ref{fig:fig3new} and \ref{fig:temperature_evolution} are inconsistent, since a phonon offset added to the (normalized) calculated electronic conductivity would significantly reduce the oscillation amplitude below the experimentally observed ${\cal O}$(1) values. 
However, we note that the observed $\sim$25\% field-angle modulation of $\kappa/T$ (c.f. Fig. \ref{fig:angle_variation}) at temperatures approaching $\sim 0.02 T_c$ implies a relatively large electronic contribution.
Outside of the unlikely possibility that phonon conductivity is strongly field- and field angle-dependent, this suggests the amplitude of $\kappa^{el}(T)$ is larger than anticipated by our simplest model. As shown in the Supplementary Information, the low-$T$ electronic contribution can indeed be increased substantially within our framework via various mechanisms described in Fig.~\ref{fig:fig3new} (Supplementary Fig.~\ref{fig:SI_Theory_3} ) or by considering two-component mixtures (Supplementary Fig.~\ref{fig:SI_Theory_1} and  \ref{fig:SI_Theory_2}), but would also require further tuning to understand higher temperature behavior.  
Beyond this, we note that we have not explored the effect of vertex corrections to $\kappa_{el}/T$ arising from extended impurity potentials or interband scattering.  While these corrections vanish in the universal low-$T$ limit of line-node superconductors \cite{Durst2000}, to our knowledge they have never been investigated in the point node case, and could potentially enhance the overall $\kappa_{el}$, without necessarily contributing significantly to the field dependence.  This hypothesis is under further investigation.

Nevertheless, all of the major features of the data that we have outlined above point to a nodal origin for these oscillatory features in $\kappa(\theta)/T$. The fact that these oscillations become stronger as we go to lower temperature clearly indicates that they are not the result of phonons, deep minima in the superconducting gap, or other ephemera. A similar conclusion follows from the very low field range in which we observe these oscillations (less than one percent of $H_{c2}$ throughout these plane of rotation.) The quasi-linearity of $\kappa/T$ with respect to $H$, the narrow minima as a function of angle and the inversion of $\kappa(\theta)/T $ at high temperatures provide further support for this basic picture. Our extensive modeling demonstrates that the location of the minima of these oscillation only makes sense if the nodes lie on or near the $b$-axis. In the absence of evidence for time-reversal symmetry breaking, this implies that the order parameter belongs to the $B_{2u}$ representation of the crystal point group, or possibly the $A_u$ representation if there happens to be an accidental node. These results represent a rare success in using the magnetic field dependence of a bulk measurement to gain direct access to a superconducting gap structure, a success that highlights the potential for more rapid progress in elucidating unconventional superconductors through similar studies in the future. \\

\section{Methods}

\noindent {\it Experimental Details:}
Field-angle-dependent data were gathered on a single crystal sample of uranium ditelluride (UTe$_2$) grown by the self-flux method \cite{Hayes2025} and previously characterized in a fixed-angle geometry experiment \cite{Hayes2025}. 
The sample, with $T_c$=2.1~K, has a residual resistivity of 0.5~$\mu \Omega$cm, yielding a residual resistivity ratio (RRR) of $\sim$ 600. 
Contacts to the sample were made with a bismuth-tin-copper solder that maintains good thermal contact throughout the temperature and field ranges studied \cite{Smith2005}.
Thermal conductivity measurements were performed by a one-heater, two-thermometer DC method in an Oxford Instruments Kelvinox 250 dilution refrigerator inserted into a two-coil vector magnet manufactured by Cryomagnetics Inc. We employed RuO$_x$ thermometers that were previously calibrated and re-calibrated in situ, which verified that the sample was in good thermal contact with the bath for all temperatures studied here. While the orientation of the a-axis was obvious from the cleave structure of the crystal, we verified the location of the $b-c$ plane by temporarily rotating the fridge ninety degrees and sweeping the field. The minimum of $\kappa/T$ aligned with the expected $b-c$ plane. 
The ability to measure thermometry with high resolution is enabled by the use of circuitry equipped with low-temperature filtering.
The overall error in the measurement is dominated by the uncertainty of the geometric factor. We estimate this error to be $\sim$25\% in the sample studied here. Statistical error arising from fluctuations in the thermometry, etc, are reflected in the scatter of the data points. 
With applied heat, a typical thermal gradient of 1-2\% of the bath temperature creates a modest increase in sample temperature above bath temperature, the average of which we use to define the sample temperature $T_\mathrm{avg}$.
In the data presented in Fig.~1, $T_\mathrm{avg}$ varies slightly between measurement sweeps, with $\sim$1-2~mK variations in comparisons between angle (fixed-field) and field (fixed-angle) sweeps that result in non-zero differences due to the steepness of the slope of $\kappa(T)/T$ near 70~mK (c.f. zero-field temperature data in Fig.~\ref{fig:SI_Theory_3}). 
To allow direct comparisons, the 100 and 250 mT angle sweep data sets have been been shifted down by 0.016 and 0.017 mW/cm-K$^2$, respectively, according to the the slope of $\kappa(T)/T$ in zero field, yielding good reproducibility between angle (fixed-field) and field (fixed-angle) sweeps.
\\

\noindent {\it Impurity Scattering Calculation: }
In the  low temperature regime ($0\lesssim \mathrm{T}\ll\mathrm{T_c}$),  the elastic scattering from impurities becomes the dominant mechanism  of the electron relaxation.  We assume that the randomly distributed impurities are pointlike and nonmagnetic in nature.  The effect of these impurities is included via self-energy calculated within self-consistent t-matrix approximation,  which includes all the scattering processes from a single impurity.  The impurity self-energy reads,
\begin{eqnarray}
\check{\hat{\Sigma}} = n_{imp} \check{\hat{\mathbb{T}}}
\end{eqnarray}
where $n_{imp}$ is the concentration of impurities,  the self-energy $\check{\hat{\Sigma}} $ and the scattering t-matrix $\check{\hat{\mathbb{T}}}$ are the 8$\times$8 matrices in the band-spin-Nambu space.  Note that the self-energy remains diagonal in the band basis, similar to the Green’s function. Furthermore, its general structure is the same as the self-energy for a single band\cite{MishraHirschfeld2023}. 
The scattering t-matrix is,
\begin{eqnarray}
\check{\hat{\mathbb{T}}} &=& \check{\hat{V}}_{imp} \frac{\check{\hat{\mathbb{1}}}}{\check{\hat{\mathbb{1}}}- \check{\hat{\mathbb{g}}} \check{\hat{V}}_{imp}},
\end{eqnarray}
where $\check{\hat{\mathbb{1}}}$ is a 8$\times$8 identity matrix,  and  $\check{\hat{V}}_{imp}$ is the impurity potential,
\begin{equation}
\check{\hat{V}}_{imp} = \begin{pmatrix} V_{11} \check{\tau}_3 & V_{12} \check{\tau}_3  \\
V_{21} \check{\tau}_3 & V_{22} \check{\tau}_3
\end{pmatrix}.
\end{equation}
Here $\check{\tau}_3=\mathrm{diag}(1,1,-1,-1)$,   $V_{11/22}$ is the intraband impurity scattering potential for the first/second band,  and $V_{12}=V_{21}$ is the interband impurity scattering potential; $\check{\hat{\mathbb{g}}}$ is the momentum integrated impurity dressed Green's function,  which includes the impurity self-energy.  Note that  the self-energy only depends on the quasiparticle energy for pointlike impurities,  and the off-diagonal self-energy \textit{i.  e.} the superconducting  order parameter does not get  renormalized by the impurity scattering,  as the impurity self-energy also remains diagonal in the Nambu space.  This occurs because the odd-parity order parameter averages to zero.  We parameterize the impurity potentials $V_{ij}\equiv\tan(\delta_s)/(\pi \sqrt{N_i N_j} )$,  where $N_{i/j}$ is the Fermi level DOS for  ${i/j}^{\mathrm{th}}$,  and $\delta_s$ controls the strength of the impurity potential.  We take $\delta_s=2$ because stronger impurities yield a finite residual thermal conductivity in the zero-temperature limit; however, even dirtier samples exhibit very small residual thermal conductivity\cite{Hayes2025}.  We further generalize the weak coupling gap equations for multiple bands \cite{MishraHirschfeld2023}.  The impurity concentration is chosen to give 0.5$\%$ suppression of the transition temperature; this value reflects the very clean nature of the sample used in field-angle resolved thermal conductivity measurements,  which has a residual resistivity ratio of a few hundred.   Once we obtain the impurity dressed Green's function,  we calculate the thermal conductivity for each band, \cite{MishraHirschfeld2023}; the total thermal conductivity is the sum of the contributions from each band.  In the presence of a magnetic field,  the Doppler shift to quasiparticle energy makes the Green's function,   the impurity self-energy,  and the thermal conductivity functions of distance from the vortex core ($r$) and the winding angle ($\Psi$).  Therefore,  we average the local thermal conductivity over the vortex unit cell.  For the heat current $j_Q \perp H$,  we perform this averaging in a series configuration \cite{Kuebert1998},
\begin{eqnarray}
\kappa^{-1} = \frac{1}{\mathcal{A}} \int_{\mathrm{unit~cell}} d\Psi dr r \frac{1}{\kappa(r,\Psi)},
\end{eqnarray}
where $\mathcal{A}$ is the vortex unit cell area.

\section{Data availability}

\bibliographystyle{naturemag}
\bibliography{references.bib}

\section{acknowledgements}

The authors thank N.P.~Butch, J.C.S.~Davis, V.~Madhavan and B.~Ramshaw for useful discussions.

\section{Funding Statement}

Research at the University of Maryland was supported by the Gordon and Betty Moore Foundation’s EPiQS Initiative Grant No. GBMF9071 (materials synthesis),
the Department of Energy, Office of Basic Energy Sciences Award No. DE-SC-0019154 (experimental measurements), 
the NIST Center for Neutron Research, and the Maryland Quantum Materials Center. 
Research at the University of Florida was supported by NSF-DMR-2231821.

\section{Author Contribution Statement}

I.M.H. and E.F. performed the thermal transport measurements. I.M.H. and S.S. synthesized the samples studied. V.M. performed the theoretical modeling. P.J.H. and J.P. supervised the project. All the authors participated in the interpretation of the data and calculations.  I.M.H., V.M., P.J.H., and J.P. wrote the manuscript.

\section{Competing Interests Statement}

The authors declare no competing financial interests.

\clearpage

\pagebreak 
\newpage 

\onecolumngrid
\begin{center}
\textbf{\large{}\textemdash{} Supplemental Information \textemdash{}}
\par\end{center}{\large \par}
\begin{center}
\textbf{\large{}Field-angle-resolved heat transport in UTe$_2$: determination of nodal positions in the superconducting order parameter}
\par\end{center}{\large \par}
\begin{center}
{Ian M. Hayes,$^{1}$ Elliot Fang,$^1$ Shanta R. Saha,$^1$ Vivek Mishra,$^{2}$ P. J.  Hirschfeld,$^{2}$ Johnpierre Paglione,$^{1,3}$}\\
{\small $^{1}$Maryland Quantum Materials Center, Department of Physics,  \\ University of Maryland, College Park, MD 20742, USA.\\
$^{2}$Department of Physics, University of Florida, Gainesville, FL, 32611-8440,  USA.\\
$^{3}$The Canadian Institute for Advanced Research, Toronto, Ontario, Canada.}
\end{center}

\vspace*{2mm}
Below we provide additional experimental data and theoretical  results supplementing the conclusions from the main text.
\setcounter{figure}{0}
\setcounter{equation}{0}
\renewcommand{\figurename}{Supplementary Fig.}
\renewcommand{\thefigure}{S\arabic{figure}}%
\setcounter{table}{0}
\renewcommand{\thetable}{A\arabic{table}}

\twocolumngrid

\renewcommand{\thefigure}{\arabic{figure}}

\section{\texorpdfstring{\boldmath{$\kappa(\theta,H)$}}{Field-Angle-Dependent Thermal Conductivity} at large magnetic fields}

\begin{figure}
 \includegraphics[width=0.95\linewidth]{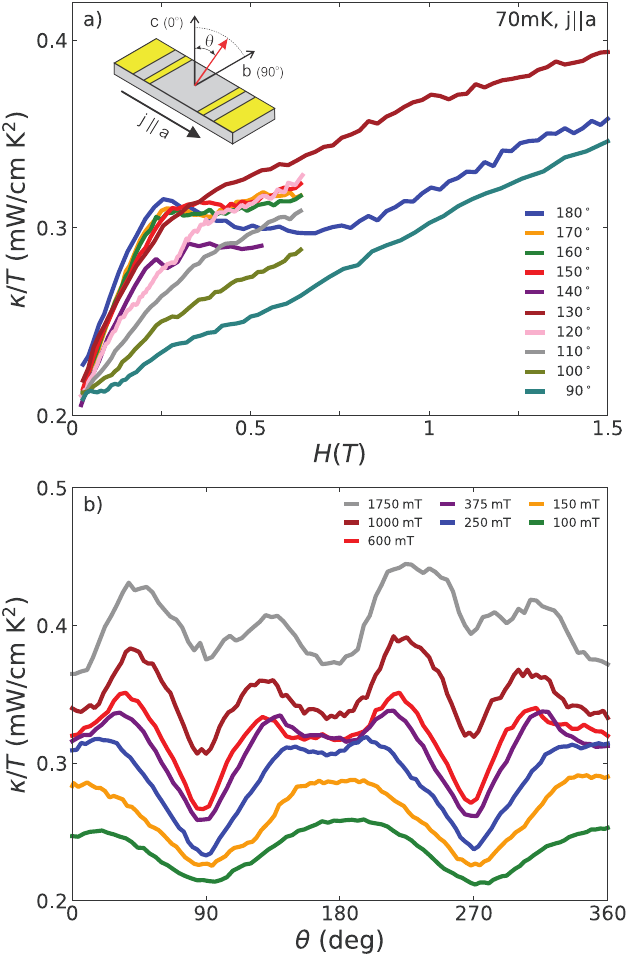} 
\caption{
{\bf Magnetic field-angle dependence of UTe$_2$ thermal conductivity up to 1750 mT.} 
Panel a) presents the fixed-angle field sweeps with heat current applied parallel to the $a$-axis and field angles rotated through the perpendicular $b$-$c$ plane.
Panel b) shows the fixed-field angle sweeps. All data are measured at a fixed temperature of 70~mK. 
} 
\label{fig:high_field} 
\end{figure}

The main body of this article focuses on the behavior of $\kappa/T(H, \theta)$ for small values of the magnetic field ($H \sim 0.02 H_{c2}$) as well as the theoretical modeling of this regime. Small values of the field are easier to handle theoretically, allowing us to reach more robust conclusions from the data collected at this field scale. { Because of the complexity of the physics in the mixed state, we don't think that any of these observations undercut our analysis of the low-field data. There is nothing like a sharp kink in these data (or any other data that we are aware of) that would suggest that the system undergoes a phase transition as a function of field in this regime. The simplest and most likely scenario is therefore that the symmetry of the superconducting state is the same throughout this field regime. However, just as with the fixed-angle configuration ($H\parallel j_Q \parallel a$) that was the subject of our prior work \cite{Hayes2025}, there are distinctive phenomena that appear in the angle-dependent $\kappa/T(H)$ at higher fields that could harbor clues to the exact nature of the superconducting state and the nature of vortex physics in this system}. Figure \ref{fig:high_field} shows a change in $\kappa/T$ with increasing field up one and a half tesla. There is a clear and qualitative change above 300-500~mT to a new oscillatory pattern as a function of the angle of the applied field. For most orientations of the applied field $\kappa/T$ increases monotonically as a function of increasing field, albeit with a clear change in slope above 300~mT. However, for a narrow range of fields near the crystal $c-$axis, $\kappa/T$ actually decreases slightly before renewing its upward climb at a slope that matches the slope of $\kappa/T(H)$ exhibited at all other angles. This creates a local maximum near 300~mT that is qualitatively very similar to the field dependence for $H\parallel a$ that we reported in our previous work.

It would be dangerous to speculate too much on the origin of this behavior at higher fields; there is limited guidance available from theory for the high-field regime, and UTe$_2$ has a remarkably intricate phase diagram as a function of magnetic field. However, some obvious and distinctive features of this behavior seem worth articulating here. First, the emergence of four distinct maxima at relatively low fields (around 6$\%$ of $H_{c2}$) suggests the possibility of more structure in the superconducting gap than considered above. These could be regions of k-space where the superconducting gap is finite but has pronounced minima, or they could indicate that there are actually multiple nodes near the crystal $b$-axis, at least at intermediate fields. This provides an additional reason to consider the two-component states discussed below (section III of the Supplementary Information). Second, in this new regime the slope of $\kappa/T(H)$ is the nearly the same independent of field angle. The entire oscisllaotry pattern seems to simply undergo a rigid shift up with increasing field. This is readily apparent in panel b of figure \ref{fig:high_field}, where the slope of three curves is very similar above half a tesla, and it is also apparent in panel a of that same figure where the curves for the 1T and 1.6T rotations are nearly identical but offset. The fact that the data appear to decompose into a field-independent oscillatory component that sits atop a linearly increasing background is strongly reminiscent of the behavior observed for this sample in the field parallel to a geometry reported in \cite{Hayes2025}. In that work we identified an analogous pattern in the behavior of $\kappa/T(H)$ at intermediate fields when comparing samples with different disorder levels. All of the samples studied in that work showed the same slope in $\kappa/T(H)$ as a function of field at high field. The effect of disorder simply seemed to be to produce a rigid shift of the curve downwards for higher disorder levels, resulting in a different extrapolated intercept at zero field. In \cite{Hayes2025} we interpreted this behavior as arising from two effectively parallel conduction channels, and the observations here lend some further weight to that interpretation.

This transition to a distinct high field regime in the angle-dependent data also sheds some light on a curious fact about our data in \cite{Hayes2025}. Of the four samples studied in that paper, only the cleanest one---which is also the subject of the present work---showed a peak in $\kappa/T(H)$ when the field was oriented along the crystal $a$-axis. Because the peak occurred at the boundary between two regimes, we concluded it was not extrinsic and most likely a manifestation of the physics that governed the transition from one regime into another. The patterns observed here strongly confirm that this peak is a part of a complex transition to a new field regime. There is a peak also for $\kappa/T(H)$ in the $H\parallel c$ configuration, but not for the rest of the $b$-$c$ plane. Because these peaks are so closely tied to the crystal axes, it is highly unlikely that this could be caused by an impurity phase or other extrinsic phenomenon.

\begin{table*}
\begin{tabular}{lccccccccccccc}
\hline
Tight binding parameters &$\mu_\mathrm{U}$ & $\Delta_\mathrm{U}$ & $t_\mathrm{U}$ & $t'_\mathrm{U}$ & $t_\mathrm{ch,U}$ & $t'_\mathrm{ch,U}$ & $t_\mathrm{z,U}$ & $\mu_\mathrm{Te}$ & $\Delta_\mathrm{Te}$ & $t_\mathrm{Te}$ &  $t_\mathrm{ch,Te}$ & $t_\mathrm{z,Te}$ &  $\delta$ \\
\hline
in (eV) & -0.355 &  0.38 & 0.17 & 0.08& 0.015& 0.01& -0.0375& -2.25& -1.4& -1.5& 0& -0.05& 0.13 \\
\hline
\end{tabular}
\caption{Tight binding parameters from Ref.  \onlinecite{ShuqiuWang2025} used to model the electronic structure of UTe$_2$.  }
\label{Tab:tb}
\end{table*}

\section{Additional results for single component states}

We examine  the temperature dependence of field-angle dependent thermal conductivity along the $a$-axis for the $A_u$,  $B_{1u}$ and $B_{3u}$ states at $H=100$mT field rotating in the $b$-$c$-plane shown in Supplementary Fig. \ref{fig:SI_Theory_4}.  For this specific calculation, we set the coefficients of the basis function to unity and considered only next-nearest neighbor pairing ($\alpha=1$).
\begin{figure*}[tpph]
 \includegraphics[width=0.95\linewidth]{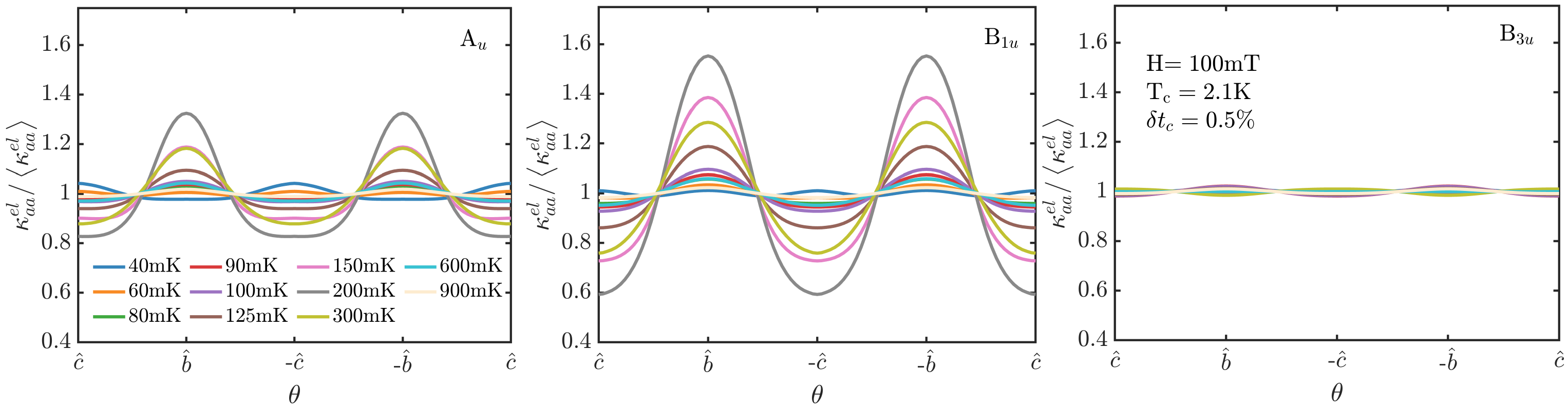} 
\caption{{\bf Variation of magnetic field-angle dependence of thermal conductivity for the $A_u$, $B_{1u}$ and $B_{3u}$ states.} Temperature evolution of field angle dependent thermal conductivity for $A_u$, $B_{1u}$ and $B_{3u}$ states at 100 mT. The transition temperature is $2.1$K for all cases and impurity concentration is chosen to give $0.5$\% suppression of $T_c$. The legends are same for all panels.  }
\label{fig:SI_Theory_4} 
\end{figure*}

Next, we investigate how various parameters of the superconducting order influence $\kappa_{el}$ at zero-field along the $a$-axis for the $B_{2u}$ state. To understand the specific trends driven by each parameter,  we isolate and vary them one at a time. The left panel of Supplementary Fig. \ref{fig:SI_Theory_3}~  illustrates the effect of unequal gaps,  $\Delta_X$ on bands centred at ($\pm \pi,/a, 0,0$) and $\Delta_Y$ on bands centred at ($0,\pm \pi/b,0$).  This variation enhances $\kappa_{el}$.  We varied the relative gap sizes while keeping the basis function coefficients at unity and restricting pairing to next-nearest neighbors.   In the temperature regime of interest,  inelastic scattering is negligible, and quasiparticle relaxation is dominated exclusively by elastic impurity scattering. The central panel shows the effect of first-neighbor and second-neighbor pairing on $\kappa_{el}$ for the $B_{2u}$ state.  In the pure next-nearest pairing limit,  nodes are located perpendicular to the heat current leading to a very small $\kappa_{el}$ contribution along the $a$-axis  ($\propto T^4$).  As the nearest neighbor pairing becomes stronger,  one pair of nodes on the Fermi pockets centered around ($\pm \pi/a,0,0$) moves towards $a$-axis,  thereby increasing $\kappa_{el}$.  Finally,  we look at the effect of basis function coefficient $c_z$ on $\kappa_{el}$ shown in the right panel of Fig. \ref{fig:SI_Theory_3}.  A smaller value of $c_z$ creates low energy excitations near $a$-axis which causes an increase in $\kappa_{el}^{aa}/T$.

\begin{figure*}
\includegraphics[width=0.95\linewidth]{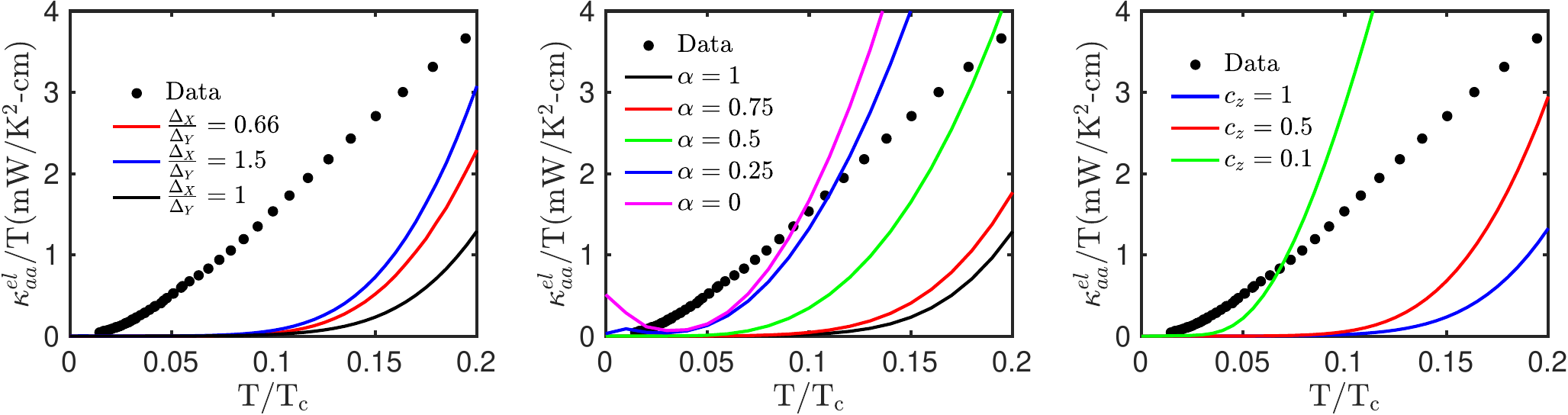} 
\caption{{\bf Low temperature zero field thermal conductivity for the $B_{2u}$ state.} Temperature dependence of thermal conductivity along the $a$-axis for the $B_{2u}$ state . The transition temperature is $2.1$ K for all cases and impurity concentration is chosen to give $0.5$\% suppression of $T_c$. Each panel shows the parameter varied in the calculations.  }
\label{fig:SI_Theory_3} 
\end{figure*}

\section{\texorpdfstring{\boldmath{$\kappa(\theta,H)$}}{Field-Angle-Dependent Thermal Conductivity} for two-component states}

We consider time reversal symmetry breaking (TRSB) superconducting states which arise from combinations of two of the four real irreducible representations  of the $D_{2h}$ point group  with a relative phase of $\pi$. The effective $\mathbf{d}$-vectors reads,
\begin{eqnarray}
\mathbf{d} =\frac{ \mathbf{d}_1 + i r \mathbf{d}_2}{\sqrt{1+r^2}}.
\end{eqnarray}
Here $\mathbf{d}_{1/2}$ represents one of the irreducible representations of the $D_{2h}$ point group.  We  introduce a real valued parameter $r$ that controls the relative strength of the two disparate components.  For the subsequent analysis, we set the coefficients of the basis functions to unity and exclude nearest neighbor pairing.  Reduction in the symmetry due to TRSB causes splitting of the quasiparticle energy spectrum,  which now becomes $E_\pm = \sqrt{\xi_\mathbf{k}^2+\Delta_\pm^2}$,  where $\xi_{\mathbf{k}}$ is the fermionic dispersion and $\Delta_\pm^2 = \Delta_0^2(|\mathbf{d}|^2\pm |\mathbf{d}\times \mathbf{d}^\ast|)$.  The mixing parameter $r$ determines the presence of zeros on the Fermi surface within the lower branch of the quasiparticle spectrum, which we term spectral nodes.  These spectral nodes are distinct from the points where the $\mathbf{d}$-vector or the order parameter vanishes; instead,  they host low-energy excitations  and mimic the effect of true nodes on low-temperature thermodynamic and transport properties.  We calculate the magnetic field-angle dependent thermal conductivity for these states at $T=40$~mK and $H=100$~mT shown in Supplementary Fig.  \ref{fig:SI_Theory_1}.  The zero-field,  low-temperature electronic thermal conductivity is illustrated separately in Supplementary Fig.  \ref{fig:SI_Theory_2}. 

\begin{figure*}
 \includegraphics[width=0.95\linewidth]{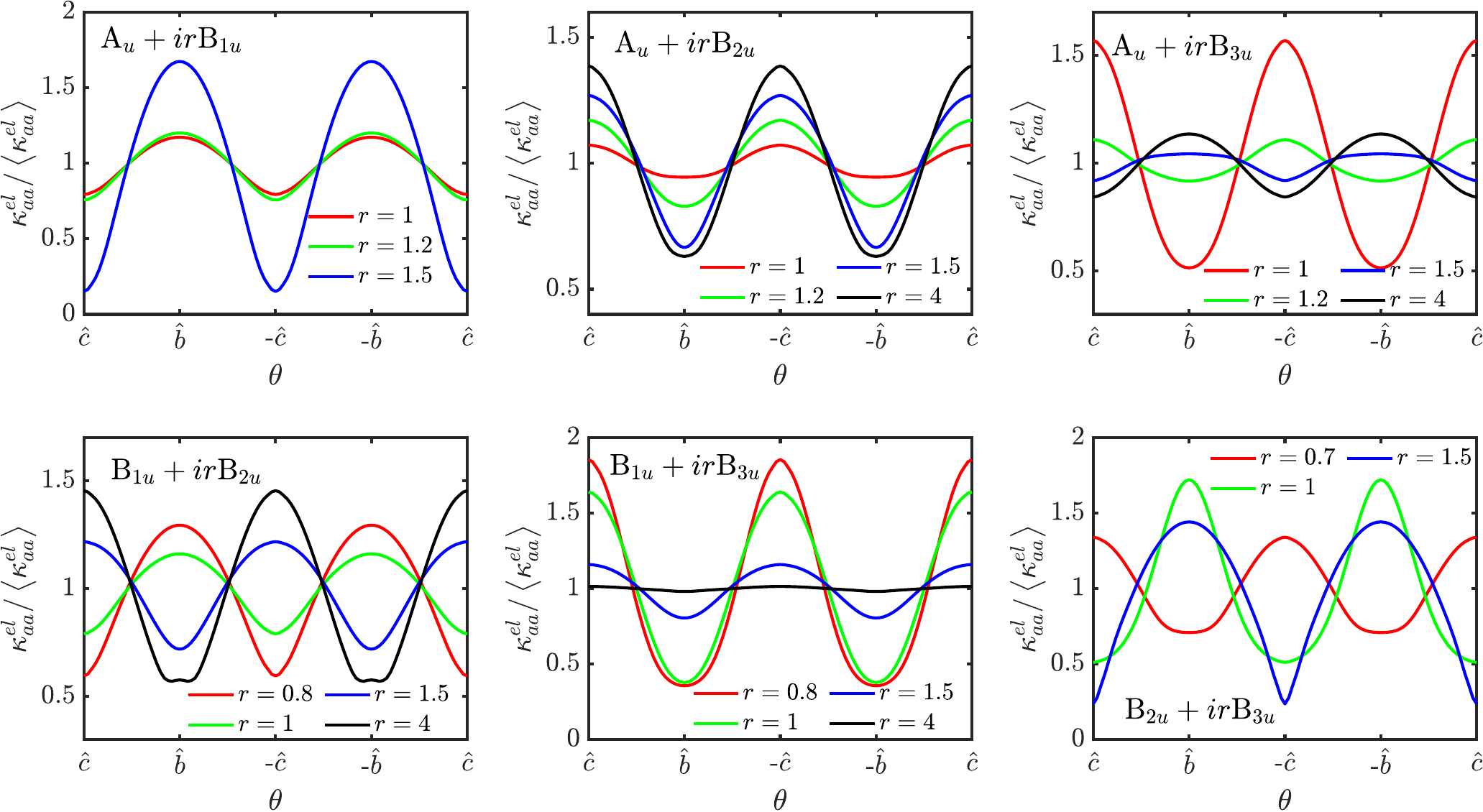} 
\caption{{\bf Magnetic field-angle dependent thermal conductivity for the two component states.} Thermal conductivity along the $a$-axis as a function of magnetic field angle $\theta$ \textit{w.r.t.} $c$-axis at 40~mK and 100~mT for all possible two component states. The transition temperature is $2.1$~K for all cases and impurity concentration is chosen to give $0.5$\% suppression of $T_c$.   }
\label{fig:SI_Theory_1} 
\end{figure*}
Only states featuring spectral nodes along the $b$-axis (or nearly so) display a minimum in the field-angle-dependent thermal conductivity when the magnetic field is aligned along that same axis. The specific states that exhibit this behavior are:
\paragraph*{${\bf A_u + i r B_{2u}} $ {\rm state}:} This state shows minima along $b$-axis when the $B_{2u}$ component is the dominant component,  positioning spectral nodes near the $b$-axis in the $ab$ and $b$-$c$ planes.  As the value of $r$ increases,  all these nodes eventually collapse to a single pair of spectral nodes oriented purely along $b$-axis in the $r\rightarrow \infty$ limit.
\paragraph*{${\bf A_u + i r B_{3u}} $ {\rm state}:} Minima appear along the $b$-axis when the components are nearly degenerate.  As the value of $r$ increases,  all these spectral nodes shift towards $a$-axis,  causing the feature along the $b$-axis to become a maximum.
\paragraph*{${\bf  B_{1u} + i r B_{2u}} $ {\rm state}:} The dominance of the $B_{2u}$ component results in spectral nodes located in closer proximity to the $b$-axis,  which leads to minima along that direction.
\paragraph*{${\bf B_{1u} + i r B_{3u}} $ {\rm state}:} When degeneracy is perfect or near-perfect,  spectral nodes align along the $b$-axis,  causing a minimum in the field-angle-dependent thermal conductivity along the $b$-axis.
\paragraph*{${\bf B_{2u} + i r B_{3u}} $ {\rm state}:} A dominant $B_{2u}$ state essential for minima along the $b$-axis.
\begin{figure*}
 \includegraphics[width=0.95\linewidth]{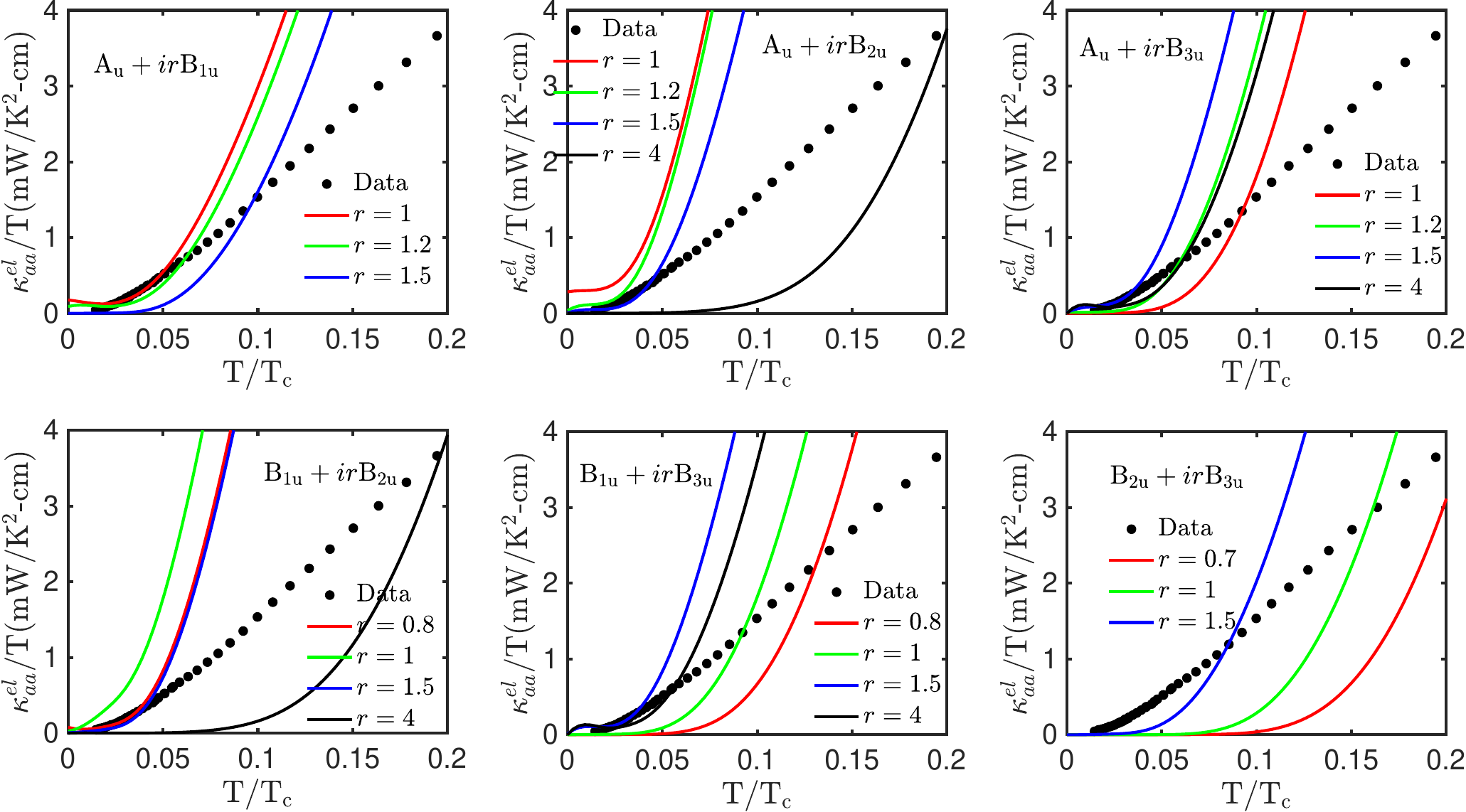} 
\caption{{\bf Low temperature zero field thermal conductivity for the two component states.} Temperature dependence of thermal conductivity along the $a$-axis for the two component states . The transition temperature is $2.1$~K for all cases and impurity concentration is chosen to give $0.5$\% suppression of $T_c$.} 
\label{fig:SI_Theory_2} 
\end{figure*}

\end{document}